\documentstyle[aps,prd,preprint,floats,epsfig]{revtex}

\newcommand{\mev}{MeV}

\newcommand{\dsone}{D_{s1}^+}

\begin{document}
\draft  

\preprint{\tighten\vbox{\hbox{\hfil CLNS 97/1513}
                        \hbox{\hfil CLEO 97-20}
}}

\title{Search For the Decay $B\to\dsone(2536) X$}

\author{CLEO Collaboration}
\date{\today}

\maketitle
\tighten

\begin{abstract} 
We have searched for the decay $B\to\dsone(2536) X$ and measured
an upper limit for the inclusive branching fraction of
${\cal B}(B\to\dsone X) < 0.95\%$ at the 90\% confidence level.
This limit is small compared with the total expected 
$B\to \bar{D}^{(*)}D^{(*)}KX$ rate. Assuming factorization, the
$\dsone$ decay constant is constrained to be 
$f_{\dsone}<114$ \mev\ at the 90\% confidence level,
at least 2.5 times smaller than that of $D_s^+$.
\end{abstract}

\newpage

{
\renewcommand{\thefootnote}{\fnsymbol{footnote}}


\begin{center}
M.~Bishai,$^{1}$ J.~Fast,$^{1}$ J.~W.~Hinson,$^{1}$
N.~Menon,$^{1}$ D.~H.~Miller,$^{1}$ E.~I.~Shibata,$^{1}$
I.~P.~J.~Shipsey,$^{1}$ M.~Yurko,$^{1}$
S.~Glenn,$^{2}$ S.~D.~Johnson,$^{2}$
% ...
\end{center}
 
\small
\begin{center}
$^{1}${Purdue University, West Lafayette, Indiana 47907}\\
$^{2}${University of Rochester, Rochester, New York 14627}\\
% ...
$^{13}${Cornell University, Ithaca, New York 14853}\\
$^{14}${University of Florida, Gainesville, Florida 32611}\\
% ...
$^{25}${University of Oklahoma, Norman, Oklahoma 73019}
\end{center}

\setcounter{footnote}{0}
}
\newpage


\section{Introduction}

One of the outstanding issues in $B$ meson physics is the semileptonic
branching fraction puzzle. Experimentally 
${\cal B}(B\to X\ell\nu)$ is measured to be ($10.43\pm0.24$)\%~\cite{PDG},
whereas theoretical calculations have difficulties accommodating a
branching fraction below $\sim 12.5\%$~\cite{Bigi}.
One way to reduce the theoretical expectations
is through a two-fold enhancement in the assumed 
$\bar{b}\to \bar{c}c\bar{s}$ rate~\cite{bccs and semilept}, 
which is estimated to be  $\sim15\%$
from the measured  inclusive rates for
$B\to D_s^+X$ and $B\to\psi X$.

\begin{figure}
\centerline{\psfig{file=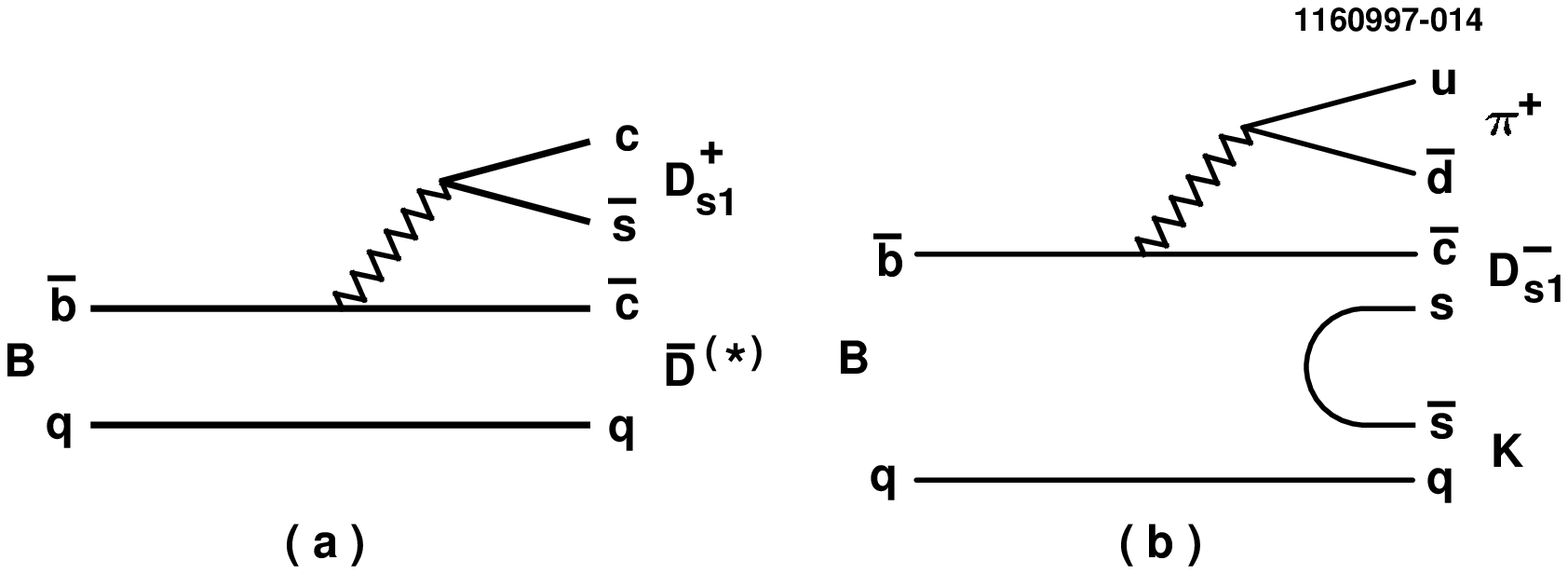,width=\textwidth}}
\medskip
\caption{Feynman diagrams for 
(a) $B\to D_{s1}^+ X$ decays producing
$D_{s1}^+$ at the upper-vertex and
(b) $B\to D_{s1}^- X$ decays producing
$D_{s1}^-$ at the lower-vertex.}
\label{fig:feynman}
\end{figure}

Recently, Buchalla  {\em et~al.}\cite{wrong D theory} and 
Blok  {\em et~al.}\cite{blok}
have suggested that a significant
fraction of the $\bar{b}\to \bar{c}c\bar{s}$ transition hadronizes into 
$B\to \bar{D}DKX$. 
This is supported by CLEO's~\cite{Dlep corr} 
observation of ``wrong-sign'' $D$ mesons from $B$ decays,
${\cal B}(B\to DX) = (7.9\pm2.2)\%$,   
where the $D$ comes from the virtual $W^+\to c\bar{s}$.
The ALEPH~\cite{ALEPH} and DELPHI~\cite{DELPHI} collaborations
have also observed sizeable
$B\to D^{(*)} \bar{D}^{(*)}X$ decay rates.
Exclusive $B$ decays involving 
wrong-sign $D$ mesons can result from
(1) resonant $B\to \bar{D}^{(*)}D_s^{**}$ decays, where
the $W^+\to c\bar{s}$ hadronizes
to an excited $D_s^+$ meson that decays
into $DKX$; and
(2) non-resonant $B\to \bar{D}^{(*)}D^{(*)}K$ decays.
This paper explores one possibility in the first case,
namely, the decays $B\to D_{s1}^+(2536) X$ 
where  $D_{s1}^+$ 
is the narrow P-wave $D_s^+$ meson with $J^P=1^+$.
The ``upper-vertex'' production of $D_{s1}^+$ from   $W^+\to c\bar{s}$
hadronization is shown in Figure~\ref{fig:feynman}(a). 
In addition, $D_{s1}^+$ mesons can be produced from ``lower-vertex''
decays $b\to c\bar{u}d$ with the creation of an $s\bar{s}$ quark pair,
as shown in Figure~\ref{fig:feynman}(b). This  produces
right-sign $D$ mesons; however, the decay rate is expected to be small.
Throughout this paper  charge conjugate states are implied.

Continuum $D_{s1}^+$ production has been thoroughly
studied~\cite{PDG}.
The $D_{s1}^+$ is just above the $D^*K$ mass threshold and decays dominantly into
$D^{*0} K^+$ and $D^{*+} K^0$.
Other possible decay channels are negligible:
$D_s^{(*)+}\pi^0$ due to isospin conservation, 
$D_s^{(*)+}(n\pi)$ due to OZI suppression~\cite{OZI}, 
$DK$ or $D^+_s\pi^0$ due to angular momentum and parity conservation, 
and $D_s^{(*)+}\gamma$ due to the small radiative decay rate.


\section{Data Sample and Event Selection}

The data used in this analysis were selected from hadronic events
collected by the CLEO~II detector at the Cornell Electron Storage Ring (CESR).
The CLEO~II detector~\cite{CLEO NIM} is a large solenoidal detector
with 67 tracking layers and a CsI electromagnetic calorimeter that
provides efficient $\pi^0$ reconstruction. 
The data consist of an integrated luminosity of 3.11 fb$^{-1}$
at the $\Upsilon(4S)$ resonance, corresponding to $3.3\times 10^6$ 
$B\bar{B}$ events. 
To evaluate non-$B\bar{B}$ backgrounds we also collected
1.61 fb$^{-1}$ of ``continuum'' data 
60 MeV below the $\Upsilon(4S)$ resonance.

The inclusive $B\to D_{s1}^+ X$ decay is studied by reconstructing
the decay channels $D_{s1}^+\to D^{*0} K^+$ and $D^{*+} K^0_S$ using
the decay modes $D^{*0}\to D^0\pi^0$ and $D^{*+}\to D^0\pi^+$. The $D^0$ is 
reconstructed using the decay modes $D^0\to K^-\pi^+$ and $K^-\pi^+\pi^0$.
Hadronic events are required to satisfy the ratio of Fox-Wolfram 
moments~\cite{R2}
$R_2=H_2/H_0<0.3$ to reduce the background from continuum events.

Charged tracks, except pions from $K^0_S$ decays, are required to
be consistent with coming from the primary interaction point.
Charged kaon and pion candidates are identified using 
specific ionization ($dE/dx$) and, when available, time-of-flight (TOF)
information.
For kaon identification, we consider
the relative probability for a charged track to be a kaon,
${\cal R}_K  = {\cal P}_K / ( {\cal P}_{\pi} + {\cal P}_K + {\cal P}_p)$,
where ${\cal P}$ is the $\chi^2$ probability for a given particle hypothesis.
The requirement on ${\cal R}_K$ depends on the decay mode of interest.
Pion candidates are identified by requiring the
$dE/dx$ and, when available,
TOF information to be within 3 standard deviations ($\sigma$) of  
that expected for pions.
We select $K^0_S$ candidates through the decay to $\pi^+\pi^-$
by requiring a decay vertex displaced from the primary interaction point
and a $K^0_S$ invariant mass within 10 MeV/c${}^2$ of its nominal value.
We reconstruct $\pi^0$ candidates through the decay to $\gamma\gamma$
by requiring candidates to have an invariant mass within 2.5 standard
deviations ($\sigma\approx 5$ MeV/c${}^2$) of the nominal $\pi^0$ mass.

The  $K^-\pi^+$ and $K^-\pi^+\pi^0$ combinations are required to have
a kaon identification of ${\cal R}_K>0.5$ and $0.7$, respectively, and
an invariant mass within 15  and 25 MeV/c${}^2$  ($\sim 2\sigma$) 
of the nominal $D^0$ mass, respectively.
In addition, we select regions of the $D^0\to K^-\pi^+\pi^0$ Dalitz plot
to take advantage of the known resonant substructure~\cite{E691}.
For the $D_{s1}^+\to D^{*0} K^+$ mode, the Dalitz cut
reduces the signal efficiency by 40\% and the background by 80\%.
We relax the Dalitz cut for the $D^{*+} K^0_S$ mode 
since the combinatoric background is substantially lower.

The  $D^{*+}\to D^0\pi^+$ candidates are
required to have a mass difference $M(D^0\pi^+)-M(D^0)$
within 1.5 MeV/c${}^2$ ($\sim 2\sigma$)
of the nominal value of 145.4 MeV/c${}^2$,
where $M(X)$ is the reconstructed invariant mass of $X$.
Similarly, the  $D^{*0}\to D^0\pi^0$ candidates are
required to have a mass difference $M(D^0\pi^0)-M(D^0)$
within 1.5 MeV/c${}^2$ ($\sim 2\sigma$)
of the nominal value of 142.1 MeV/c${}^2$. 
To form $D_{s1}^+$ candidates
charged kaons are combined with $D^{*0}$ candidates
and $K^0_S$'s are combined with $D^{*+}$ candidates.
Since the primary kaons from  $D_{s1}^+\to D^{*0} K^+$ decays
have low momentum, we can impose a stringent ${\cal R}_K>0.9$ 
requirement on the $K^+$  with negligible loss of
efficiency. The $D_{s1}^+$ candidates are required to have a scaled momentum
$x_p=p_{D_{s1}^+}/\sqrt{E^2_{beam}-M^2_{D_{s1}^+}}<0.45$, 
which is the kinematic limit for $B\to D_{s1}^+ X$ decays. 
(We ignore the negligible contributions from $b\to u$ decays.)
Upper-vertex $D_{s1}^+$ production results in
a maximum $x_p$ of 0.35, and this requirement is imposed
when determining the $D_{s1}^+$ decay constant.
The $D_{s1}^+$ decay channels with  $\pi^0$'s in the final state often
have multiple $D_{s1}^+$ candidates per event. We select the candidate
with the highest $\chi^2$ probability of being a $D_{s1}^+$, which is
derived from the invariant
masses of the reconstructed $\pi^0$, $D^0$ and $D^*$ mesons.

\section{Raw Yields}

\begin{figure}
\centerline{\psfig{file=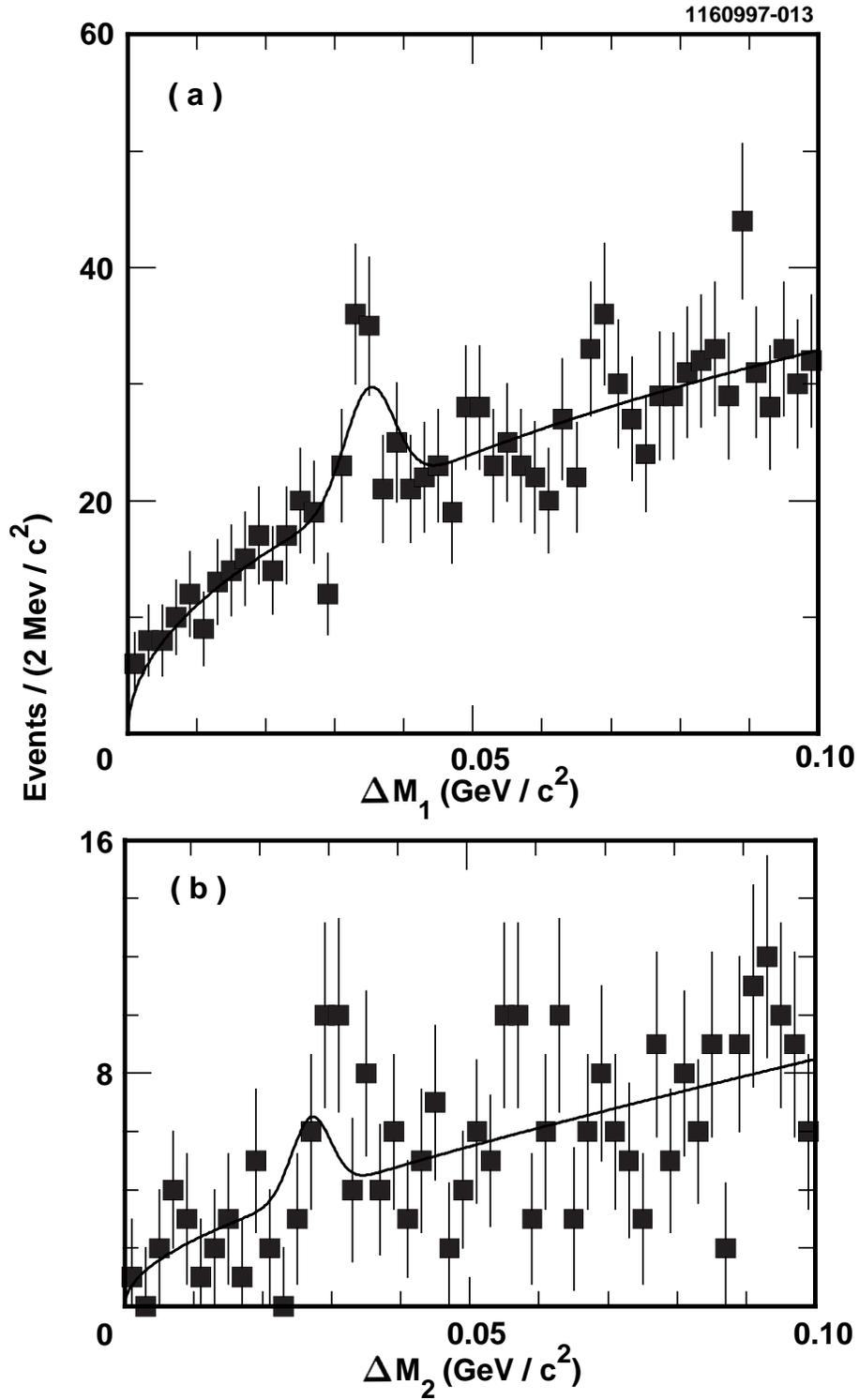,width=0.75\textwidth}}
\medskip
\caption{The mass difference distribution for (a) $D^{*0} K^+$ and
(b) $D^{*+} K^0_S$ candidates from $B$ meson decays.}
\label{fig:mass diff}
\end{figure}

The $D_{s1}^+$ signal is identified using the $D^*K$ mass 
difference, 
$   \Delta M_1 = M( D^{*0} K^+) - M( D^{*0}) - M_{K^+}   $ and
$   \Delta M_2 = M( D^{*+} K^0_S) - M( D^{*+}) - M_{K^0_S}  $,
where $M_{K^+}$ and $M_{K^0_S}$ are the known masses~\cite{PDG}.
The $D^*K$ mass difference signal has a resolution that is  
two to four times smaller than the corresponding signal in the
reconstructed $D^*K$ invariant mass distribution.
The  $\Delta M_1$ and  $\Delta M_2$  distributions
are shown in Figure~\ref{fig:mass diff}, where the  $D^0\to K^-\pi^+$ 
and $K^-\pi^+\pi^0$ modes have been added together.
The data is fit with a 
Gaussian signal and a threshold background function. The Gaussian width
is fixed to that expected from a GEANT-based Monte Carlo
simulation~\cite{geant}
($\sigma= 2.4-3.6$ MeV/c${}^2$, depending on the mode) and the mean is fixed 
to  the measured $D_{s1}^+$ mass difference
from continuum data
($\Delta M_1\approx 35$ MeV/c${}^2$ and 
$\Delta M_2\approx 27$ MeV/c${}^2$.)
We observe $42\pm14$ signal events in the $D^{*0} K^+$ mode
and $9\pm6$ events in the $D^{*+} K^0_S$ mode.

\begin{figure}
\centerline{\psfig{file=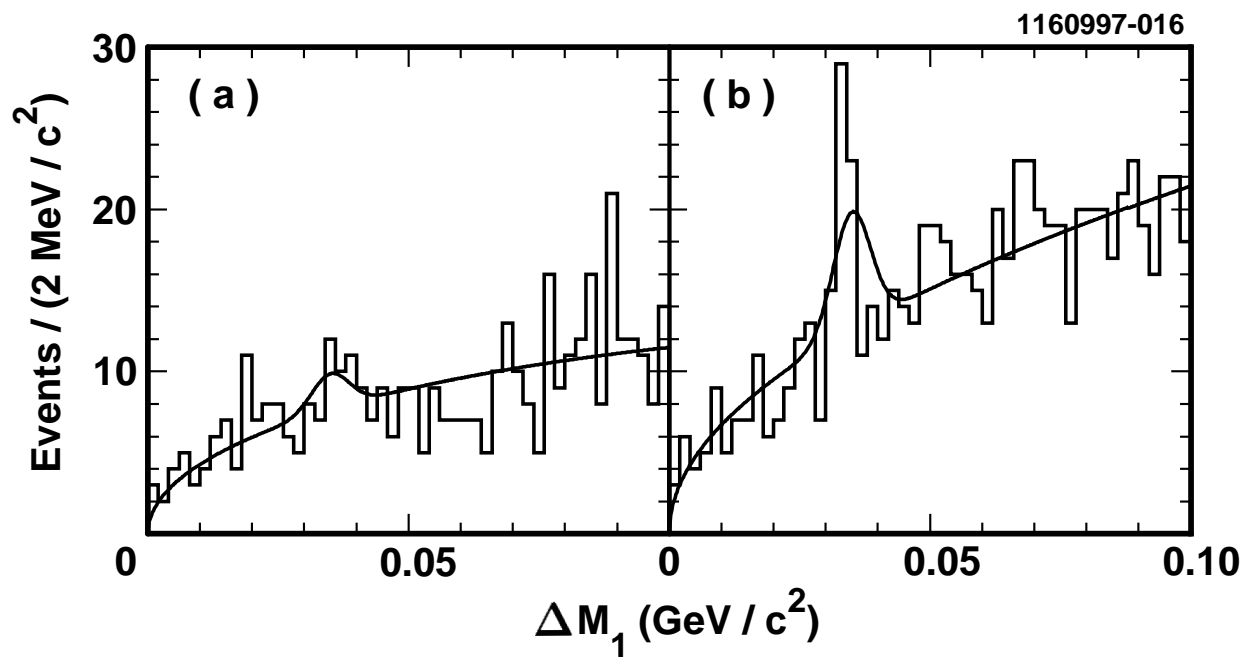,width=\textwidth}}
\medskip
\caption{The $\Delta M_1$ mass difference distribution for $D^{*0} K^+$ 
candidates from the (a) $D^0\to K^-\pi^+$ and
(b)  $D^0\to K^-\pi^+\pi^0$ decay channels.}
\label{fig:2 D0 modes}
\end{figure}

However, when the $D^{*0} K^+$ candidates  are further subdivided
into the $D^0\to K^-\pi^+$ and $K^-\pi^+\pi^0$ decay channels
there is a discrepancy
in the $D_{s1}^+$ yields. As shown in 
Figure~\ref{fig:2 D0 modes}, we observe $10\pm8$ signal events
in the $\Delta M_1$ distribution
for the $D^0\to K^-\pi^+$ channel and $33\pm12$ $D_{s1}^+$ signal events
for the $D^0\to K^-\pi^+\pi^0$ channel. After accounting for
branching fractions and efficiencies, discussed below,
this results in a $2.2\sigma$ discrepancy in the
$D^{*0} K^+$ rates between the two $D^0$ modes.
We cannot rule out the fact that background sources may be
contributing a false $D_{s1}^+$ signal in the $D^0\to K^-\pi^+\pi^0$ channel, but not
in the $D^0\to K^-\pi^+$ channel. However, no
such mechanism has been uncovered. To be conservative, we choose to quote
only an upper limit for the decay $B\to D_{s1}^+ X$.


Since the $D_{s1}^+$ reconstruction efficiency increases rapidly with $x_p$ and the
$D_{s1}^+$ momentum distribution from $B$ decays is not known, we compute the
inclusive $B\to D_{s1}^+ X$ branching fraction by dividing the
data into four equal regions of $x_p$  from 0.05 to 0.45 and summing the 
efficiency corrected yields. The $D_{s1}^+\to D^{*0} K^+$ and $D^{*+} K^0$
branching fractions are equal according to isospin, and their ratio
has been measured to be within 30\% of unity~\cite{ds1}.
We measure the branching fraction 
$B\to D_{s1}^+ X$ to be $(0.77\pm0.22)\%$ from the $D^{*0} K^+$ mode
and  $(0.28\pm0.37)\%$ from the $D^{*+} K^0_S$ mode, where the error
is statistical only. The two measurements are statistically
consistent. The  $x_p$ distribution for our $D_{s1}^+$ candidates is
shown in Figure~\ref{fig:xp dist}. 

\section{Cross-Checks}

\begin{figure}
\centerline{\psfig{file=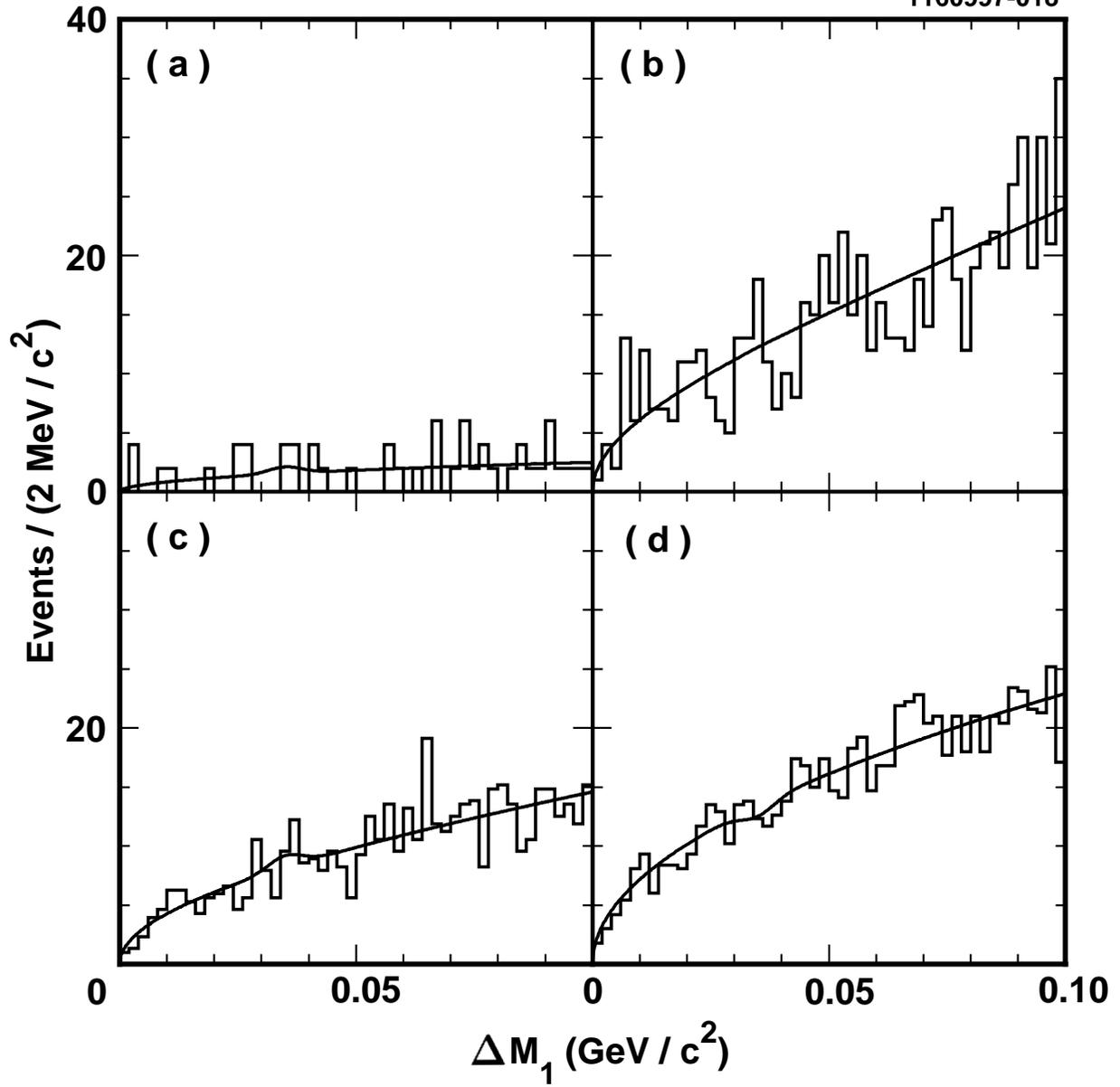,width=\textwidth}}
\medskip
\caption{The normalized $D^{*0} K^+$ mass difference distributions from
(a) continuum events, 
(b) $D^{*0} K^-$ ``wrong-sign'' combinations,
(c) $D^0$ mass sidebands, and
(d) $D^{*0}$ mass sidebands.}
\label{fig:cross checks}
\end{figure}

Several cross-checks, shown in Figure~\ref{fig:cross checks},
were performed to corroborate the validity of
the $D_{s1}^+$ signal.
The scaled continuum background  from data
after satisfying all selection cuts 
is negligible, and there is no excess in the $\Delta M_1$ signal region
($3\pm5$ events). The uncertainty in the continuum $D_{s1}^+$ contribution
is included in the  systematic error.
There is also
no evidence of peaking in the $\Delta M_1$ signal region for
wrong-sign $D^{*0} K^-$ combinations ($0\pm9$ events),
$D^0$ mass sidebands ($5\pm5$ events),
%
%
and $D^{*0}$ mass sidebands ($-4\pm6$ events).

\begin{figure}
\centerline{\psfig{file=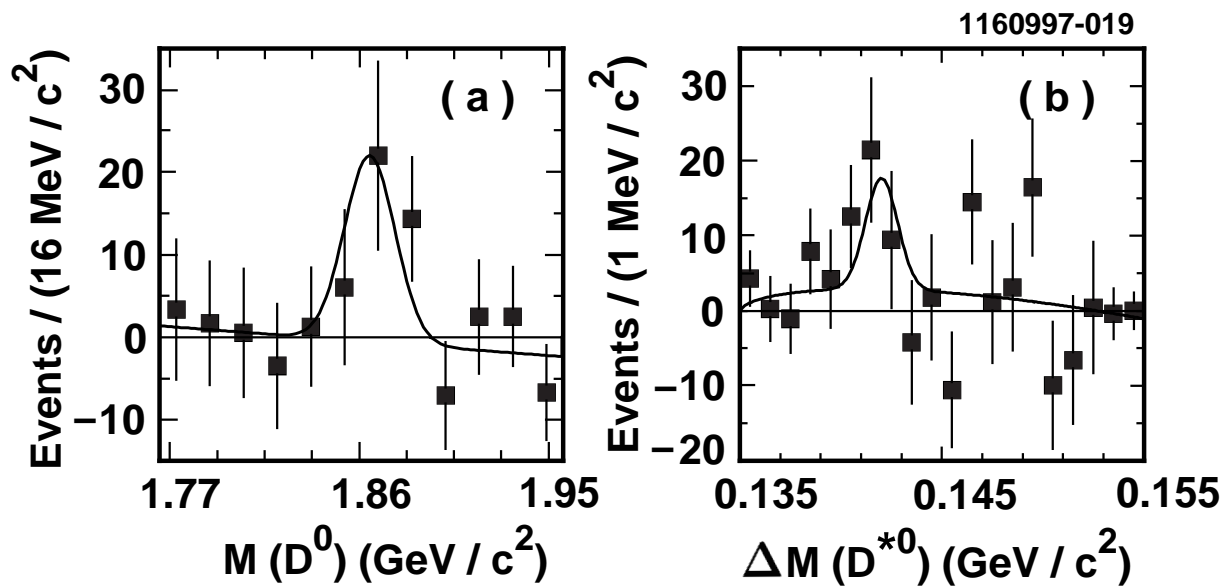,width=\textwidth}}
\medskip
\caption{(a) The invariant mass distribution for $K^-\pi^+$ and $K^-\pi^+\pi^0$ 
combinations from $D^{*0} K^+$ candidates in the $\Delta M_1$ signal 
region, after sideband subtraction.
(b) The $D^{*0}$ mass difference distribution
from $D^{*0} K^+$ candidates in the $\Delta M_1$ signal 
region, after sideband subtraction.}
\label{fig:D0 yield}
\end{figure}

We have also searched for the $D^0$ signal from $D_{s1}^+\to  D^{*0} K^+$ 
candidates in the $\Delta M_1$  signal region, 
$|\Delta M_1 - 35$ MeV/c${}^2 |<10$ MeV/c${}^2$,
by relaxing the $D^0$ mass cut and histogramming the invariant mass 
of all $K^-\pi^+$ and $K^-\pi^+\pi^0$ combinations that satisfy
the remaining selection criteria. 
In events with multiple candidates per $D^0$ decay mode we select the
candidate with the highest $\chi^2$ probability, 
which is derived from the
reconstructed $\pi^0$ and $D_{s1}^+$  masses.
We observe $100\pm15$ $D^0$ events. 
However, 
there are also real $D^0$'s in the random $D^{*0} K^+$ combinations
under the $D_{s1}^+$ peak;  
after a $\Delta M_1$
sideband subtraction  the $D^0$ invariant mass spectrum yields 
$44\pm18$ events (see Figure~\ref{fig:D0 yield}(a)).
This is consistent with our $D_{s1}^+\to D^{*0} K^+$ yield in 
Figure~\ref{fig:mass diff}.

Similarly, we have studied the $D^{*0}$ signal from $D_{s1}^+\to  D^{*0} K^+$ 
candidates in the $\Delta M_1$  signal region. 
We observe $59\pm15$ $D^0$ events. 
As in the $D^0$ case
there are also real $D^{*0}$'s in the random $D^{*0} K^+$ combinations
under the $D_{s1}^+$ peak.  
After a $\Delta M_1$
sideband subtraction  the $D^{*0}$ mass difference spectrum yields 
$25\pm18$ events (See Figure~\ref{fig:D0 yield}(b)), 
consistent with our $D_{s1}^+\to D^{*0} K^+$ yield.
%

\begin{figure}
\centerline{\psfig{file=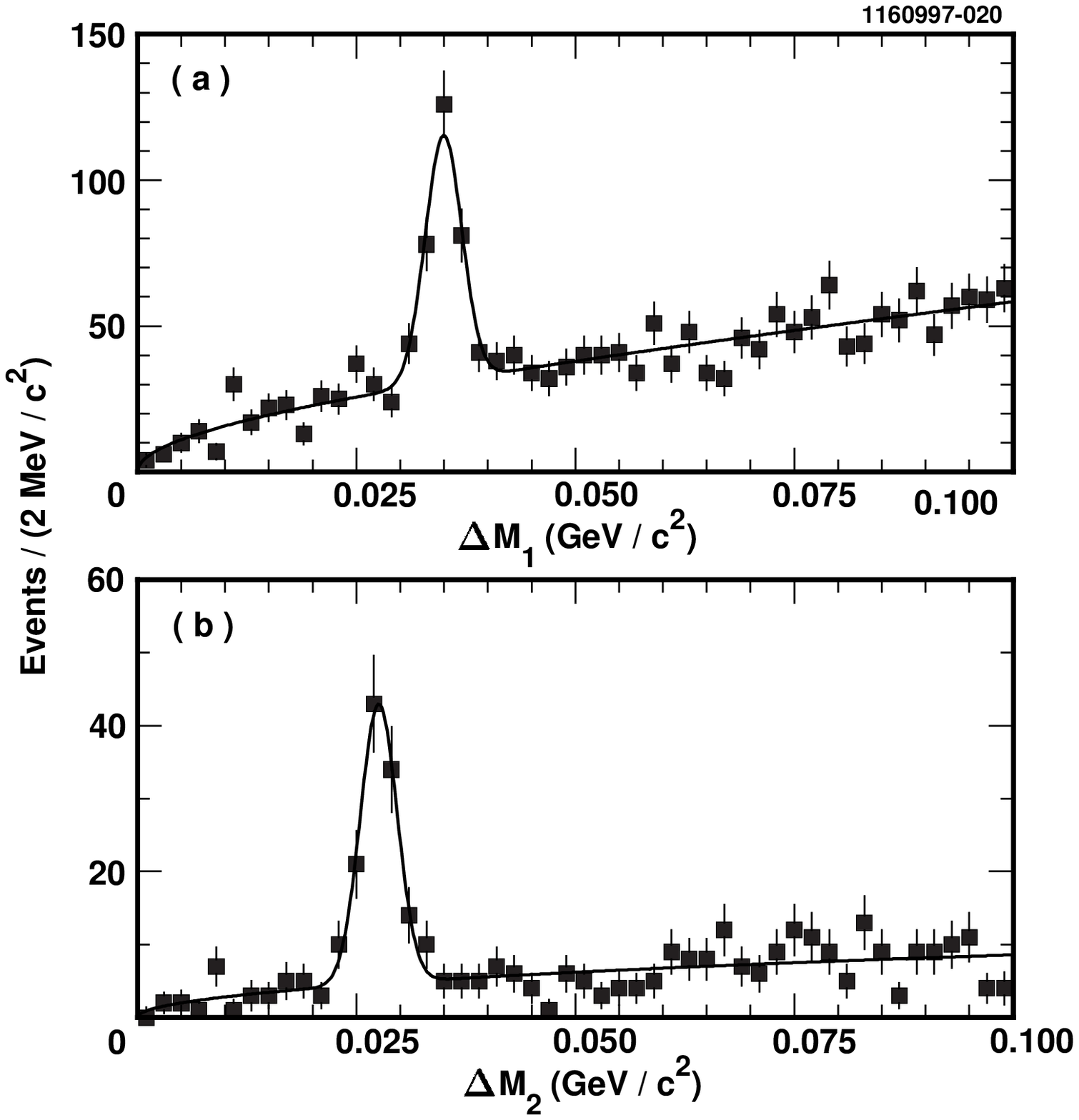,width=1.0\textwidth}}
\medskip
\caption{The mass difference distribution for (a) $D^{*0} K^+$ and
(b) $D^{*+} K^0_S$ candidates from continuum $e^+e^-\to c\bar{c}$ 
events.}
\label{fig:continuum}
\end{figure}

Finally, we have studied the $D_{s1}^+$ production  from continuum
$e^+e^-\to c\bar{c}$ events.
The selection criteria is similar to that used to
find $D_{s1}^+$ from $B$ decays, but since continuum charm 
production has a hard 
fragmentation, we require $x_p>0.5$. In addition, we remove the $R_2<0.3$ cut,
relax the charged kaon identification to ${\cal R}_K>0.1$, 
and remove the Dalitz cut for $D^0\to K^-\pi^+\pi^0$.
The mass difference distribution for  $D^{*0} K^+$ and
$D^{*+} K^0_S$ combinations are shown in Figure~\ref{fig:continuum},
where the $D^0\to K^-\pi^+$ and $K^-\pi^+\pi^0$ modes have been added together.
We extract the $D_{s1}^+$ signal by fitting the data with a Gaussian signal
and a threshold background function. The Gaussian width is fixed to the
value predicted by Monte Carlo (2.1 MeV/c${}^2$), and the mean is allowed to float.
We observe $222\pm19$ events in the $D_{s1}^+\to D^{*0} K^+$ mode with a
mass difference of $35.0\pm0.2$  MeV/c${}^2$ (statistical error only),
and $101\pm11$ events  in the $D_{s1}^+\to D^{*+} K^0_S$ mode with a
mass difference of $27.5\pm0.3$ MeV/c${}^2$.
The results are consistent with the previous CLEO analysis~\cite{ds1}.

\section{Systematic Errors and Final Results}

There are several sources of systematic error. 
We assign a systematic
error of 16\% to account for the $2.2\sigma$ discrepancy between the
$D^{*0} K^+$ rates for the $D^0\to K^-\pi^+$ and $K^-\pi^+\pi^0$ modes.
This accomodates different methods of computing the weighted average
of the $B\to D_{s1}^+ X$ branching fraction 
from the four separate decay chains.
Uncertainties due to reconstruction efficiencies include
1.5\% per charged track, 5\% per $\pi^0$, 5\% for slow pions from $D^*$,
and 5\% for $K^0_S$. 
We also include systematic errors of 7\% for 
Monte Carlo statistics, 5\% for kaon identification
and the Dalitz decay cut efficiency, 4\% for uncertainties in the
yield for $x_p<0.05$, and 8\% for uncertainties in the continuum
$D_{s1}^+$ contribution that passes our selection criteria.
The total systematic error is 24\%. 

Averaging the $D^{*0} K^+$ and
$D^{*+} K^0_S$ modes together, we obtain
${\cal B}(B\to D_{s1}^+ X) = (0.64\pm0.19\pm0.15)\%$.
Since the $D_{s1}^+$ signal is observed largely in only one decay mode
$D_{s1}^+\to D^{*0} K^+$ with $D^0\to K^-\pi^+\pi^0$, and since there is a
discrepancy between this mode and the corresponding mode
involving $D^0\to K^-\pi^+$,
we instead prefer to quote an upper limit on the branching
fraction to be ${\cal B}<0.95\%$ at the 90\% C.L.~\cite{upperlimit}
This decay rate limit is small relative to the total rate expected for 
$B\to \bar{D}^{(*)}D^{(*)}KX$ of about $(7.9\pm2.2)\%$ 
from the wrong-sign $D$ meson yield in $B$ decays~\cite{Dlep corr}.
This is not surprising considering  the $c\bar{s}$ system has
appreciable phase space beyond the $D_{s1}^+$ mass~\cite{wrong D theory}.
Also, CLEO's~\cite{Briere} recent observation of exclusive 
$B\to \bar{D}^{(*)}D^{(*)}K$ decays shows that the  $D^{(*)}K$
invariant mass distribution lies mostly above the $D_{s1}^+$ mass.


\section{$D_{s1}^+$ Decay Constant}

Measurement of the $B\to D_{s1}^+ X$ decay rate 
also provides an
estimate of the $D_{s1}^+$ decay constant, $f_{D_{s1}^+}$,
assuming that the $D_{s1}^+$ comes dominantly from upper-vertex decays. 
The inclusive decay rate for $B$ mesons into ground state or excited
$D_s^+$ mesons can be calculated assuming factorization~\cite{hitoshi},
$$  \Gamma(B\to D_s X) = {G_F^2 |V_{cb}V_{cs}|^2 \over 16\pi} 
			   M_b^3 a_1^2 f_{D_s}^2 I(x,y) $$
where  $a_1$ is the BSW~\cite{BSW} parameter for the effective 
charged current, and $I(x,y)$ is a
kinematic factor with $x=M_{D_s}^2/M_b^2$ and $y=M_c^2/M_b^2$. 
For scalar or pseudoscalar $D_s$ mesons, 
$I(x,y)= \sqrt{(1-x-y)^2-4xy}(1-x-2y-xy+y^2)$, and for vector or
axial-vector $D_s$ mesons, $I(x,y)=\sqrt{(1-x-y)^2-4xy}(1+x-2x^2-2y+xy+y^2)$.

\begin{figure}
\centerline{\psfig{file=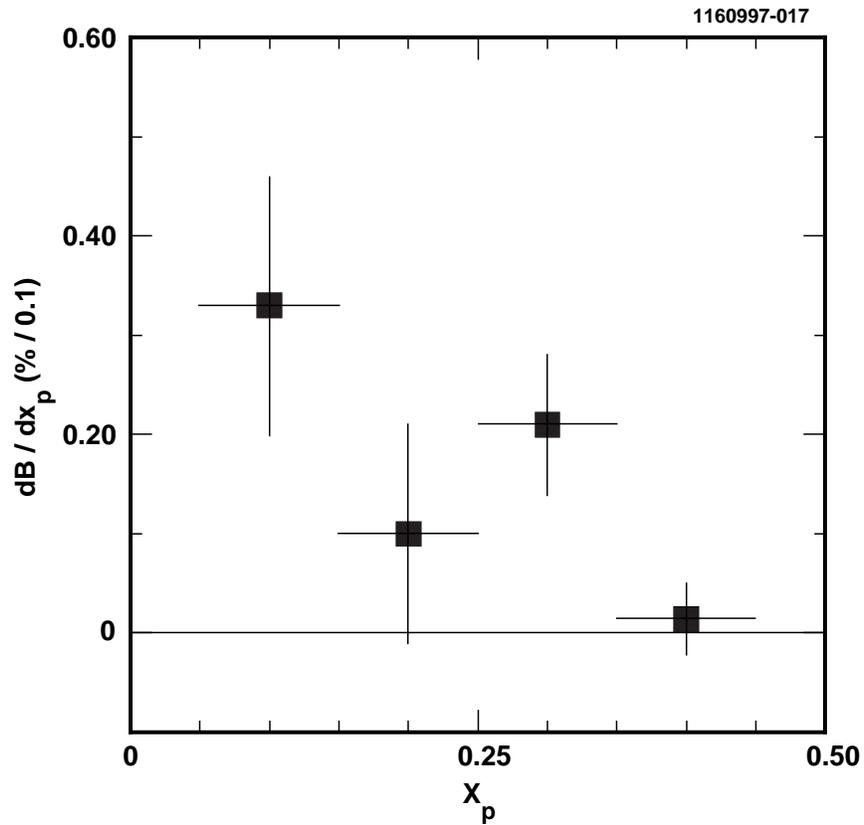,width=0.7\textwidth}}
\medskip
\caption{The efficiency corrected yield 
for our $B\to D_{s1}^+ X$ candidates as a function of
the $D_{s1}^+$ scaled momentum $x_p$. 
The kinematic limit from 
upper-vertex and lower-vertex $B\to D_{s1}^+ X$ decays is 
$x_p<0.35$ and $x_p<0.45$, respectively.}
\label{fig:xp dist}
\end{figure}

We have tightened the $x_p$ requirement to $x_p<0.35$ since this
is the kinematic limit for upper-vertex $B\to D_{s1}^+\bar{D}X$ decays.
The production of  ground state and excited 
$D_s^+$ mesons from lower-vertex decays such as $\bar{B}\to D_{s1}^+\bar{K}X$
is expected to be suppressed. 
This is certainly true for $B\to D_s^+ X$ decays where the
fraction of $D_s^+$ produced at the lower-vertex is measured to be
$0.172\pm0.079\pm0.026$~\cite{Ds lepton}.
Moreover, there is no evidence of $D_{s1}^+$ production in the region 
$x_p=0.35-0.45$ where lower-vertex production is likely
to occur (see Figure~\ref{fig:xp dist}.)

With the assumption  $f_{D_s^+} = f_{D_s^{*+}}$ we can extract
$f_{D_{s1}^+}$ from the ratio of inclusive rates, 
$$ {{\cal B}(B\to D_{s1}^+ X) \over {\cal B}(B\to D_s^+ X)}
   = {\Gamma(B\to D_{s1}^+ X) \over \Gamma(B\to D_s^+ X) + \Gamma(B\to D_s^{*+}X)}
   \approx 0.49 \left({f_{D_{s1}^+} \over f_{D_s^+}} \right)^2 $$
Many systematic errors cancel in the ratio.
When computing the $D_{s1}^+$ decay constant from the above equation,
we use $(75\pm25)\%$ of the measured $B\to D_{s1}^+ X$ branching fraction
to account for uncertainties in the upper and lower vertex contributions
to $D_{s1}^+$. This accomodates the excess of
$B\to D_{s1}^+ X$ candidates observed at low $x_p<0.15$ 
as seen in Figure~\ref{fig:xp dist}.
From our upper limit on $B\to D_{s1}^+ X$ and CLEO's~\cite{B to Ds}
measurement of
${\cal B}(B\to D_s^+ X) = (12.11\pm0.39\pm0.88\pm1.38)\%$,
we derive
$f_{D_{s1}^+} / f_{D_s^+}<0.40$ at the 90\% C.L. The central value is
$f_{D_{s1}^+} / f_{D_s^+}=
0.29\pm0.06\pm0.06$, where the first error
is due to the total error in the inclusive $B\to D^+_s X$ and 
$B\to D_{s1}^+ X$ 
branching fractions, and the
second is  the uncertainty in the
non-factorizable  and lower-vertex contributions to the 
$B\to D_{s1}^+ X$ decay rate.
Using the measured value of $f_{D_s^+}=280\pm40$ MeV ~\cite{B to Ds} 
gives $f_{D_{s1}^+}=81\pm26$ MeV which corresponds to an upper limit
of  $f_{D_{s1}^+}<114$ MeV.
This limit accomodates
the prediction of  $f_{D_{s1}^+}=87\pm19$ MeV by
Veseli and Dunietz~\cite{f_ds1}. 

\section{Conclusions}

In summary, we have searched for $B$ mesons decaying into the P-wave
$D_{s1}^+(2536)$ meson. The upper limit of
${\cal B}(B\to D_{s1}^+ X)<0.95\%$ at the 90\% C.L.
accounts for at most 
only a fraction of the total wrong-sign $B\to DX$ rate. 
Assuming factorization,
the decay constant $f_{D_{s1}^+}$ is at least
a factor of 2.5 times smaller than the
decay constant for the pseudoscalar $D_s^+$.

\bigskip
\centerline{\bf ACKNOWLEDGEMENTS}
\smallskip
We gratefully acknowledge the effort of the CESR staff in providing us with
excellent luminosity and running conditions.
J.P.A., J.R.P., and I.P.J.S. thank                                           
the NYI program of the NSF, 
M.S. thanks the PFF program of the NSF,
G.E. thanks the Heisenberg Foundation, 
%
%
K.K.G., M.S., H.N.N., T.S., and H.Y. thank the
OJI program of DOE, 
J.R.P., K.H., M.S. and V.S. thank the A.P. Sloan Foundation,
R.W. thanks the 
Alexander von Humboldt Stiftung, 
M.S. thanks Research Corporation, 
and S.D. thanks the Swiss National Science Foundation 
for support.
This work was supported by the National Science Foundation, the
U.S. Department of Energy, and the Natural Sciences and Engineering Research 
Council of Canada.


\newpage

\begin{thebibliography}{99}
\bibitem{upperlimit}
The 90\% upper limit is derived, assuming Gaussian statistics,
by adding 1.28 times the total error to the central value.
\end{thebibliography}
\end{document}